\documentclass[12pt]{article}
\usepackage{epsf}
\usepackage{amsmath,amssymb}
\usepackage{latexsym}
\usepackage{graphicx,color}
\usepackage{wrapfig}
\usepackage{latexsym}
\usepackage{graphics}
\usepackage{cite}

\usepackage{color}
\usepackage{delarray,color}
\usepackage{fancybox}  
\newcommand {\beq}{\begin{eqnarray}}
\newcommand {\eeq}{\end{eqnarray}}

\usepackage{tabularx}
\usepackage{cite}

\catcode`\@=11
\@addtoreset{equation}{section}

\catcode`@=12
\relax

\newcommand{\bCx}{\mathbb{C}^{\times}}

\newcommand{\bC}{\ensuremath{\mathbb{C}}}

\newcommand{\bP}{\ensuremath{\mathbb{P}}}

\newcommand{\bR}{\ensuremath{\mathbb{R}}}

\newcommand{\bT}{\ensuremath{\mathbb{T}}}

\newcommand{\bZ}{\ensuremath{\mathbb{Z}}}

\newcommand{\scA}{\ensuremath{\mathcal{A}}}

\newcommand{\scD}{\ensuremath{\mathcal{D}}}

\newcommand{\scM}{\ensuremath{\mathcal{M}}}
\newcommand{\scN}{\ensuremath{\mathcal{N}}}
\newcommand{\scO}{\ensuremath{\mathcal{O}}}
\newcommand{\scP}{\ensuremath{\mathcal{P}}}

\newcommand{\scV}{\ensuremath{\mathcal{V}}}

\newcommand{\scZ}{\ensuremath{\mathcal{Z}}}

\begin{document}
%%%%%%%%%%%%%%%%%%%%%%%%%%%%%%%%%%%%%%%%%%%%%%%%%%%%%%%%%%%%%%%%%%%%%%%%
\baselineskip 0.7cm

\begin{titlepage}

%% Set the number of the title with 0
\setcounter{page}{0}

%% change the footnote symbol
\renewcommand{\thefootnote}{\fnsymbol{footnote}}

\begin{flushright}
%{\tt 
UT-08-27\\
IPMU-08-0062\\
August, 2008
%\\}

\end{flushright}

\vskip 1.35cm

\begin{center}
{\Large \bf
Toric Calabi-Yau four-folds dual to Chern-Simons-matter theories
%Gravity duals of Chern-Simons-matter theories
}

\vskip 1.2cm 

{\normalsize
Kazushi U{\sc eda}$^1$\footnote{ kazushi(at)math.sci.osaka-u.ac.jp}  and Masahito Y{\sc amazaki}$^{2}$\footnote{yamazaki(at)hep-th.phys.s.u-tokyo.ac.jp}
}

\vskip 0.8cm

{ \it
$^1$ Department of Mathematics, Graduate School of Science, Osaka University, Machikaneyama 1-1, Toyonaka, Osaka, 560-0043, Japan.
\\[2mm]
$^2$Department of Physics, University of Tokyo, Hongo 7-3-1, Tokyo, 113-0033, Japan
}

\end{center}

\vspace{12mm}

\centerline{{\bf Abstract}}
We propose a new method to find gravity duals to a large class of three-dimensional Chern-Simons-matter theories, using techniques from dimer models.
 The gravity dual is given by M-theory on $AdS_4\times Y_7$, where $Y_7$ is an arbitrary seven-dimensional toric Sasaki-Einstein manifold. 
 The cone of $Y_7$ is a toric Calabi-Yau 4-fold, which coincides with a branch of the vacuum moduli space of Chern-Simons-matter theories.
% recently studied in arXiv:0808.0912. 

\end{titlepage}

\newpage

\section{Introduction}
Recently, there has been considerable developments in understanding duality between four-dimensional $\mathcal{N}=1$ superconformal quiver gauge theories and type IIB string theory on $AdS_5\times Y_5$, where $Y_5$ is a toric Sasaki-Einstein manifold. In particular, using techniques from dimer models (the so-called brane tilings \cite{BT,BT2,BT3}), we can determine toric Calabi-Yau manifolds (which is a cone of $Y_5$) dual to quiver gauge theories, and vice versa. This is a fascinating development, since it makes it possible to check AdS/CFT for infinitely many examples, or for arbitrary toric Calabi-Yau 3-folds \cite{BT3,equiv}, using for example a-maximization \cite{IW} in gauge theory side and volume minimization in gravity side \cite{MSY}.

It is thus natural to consider generalization of all these developements to the case of duality between three-dimensional conformal field theories and M-theory on $AdS_4\times Y_7$, where $Y_7$ is now a seven-dimensional toric Sasaki-Einstein manifold. This is not a mere generalization of pure academic interest: First, $AdS_4/CFT_3$ correspondence should tell us much about the landscape of four-dimensional vacua in string theory with negative cosmological constant, which could be lifted to de Sitter space by various ways (e.g. as in \cite{KKLT}). Second, some of three-dimensional supersymmetric conformal field theories have possible relevance to interesting condensed matter systems. However, this natural generalization has long been a challenging problem, since we did not understand the gauge theory side of this duality; the theory on multiple M2-branes was not known.

This situation changed dramatically quite recently. Bagger, Lambert and Gustavsson \cite{BLG} proposed a Lagrangian description of the world-volume theory of multiple M2-branes, and their theory is later generalized by \cite{ABJM}. The theory in \cite{ABJM} (the so-called ABJM theory) is 
a kind of Chern-Simons-matter theory, and is believed to be realized on the worldvolume of M2-branes probing $AdS_4\times S^7/Z_k$, where $k$ is the level of Chern-Simons term.

From our experience in the case of $AdS_5/CFT_4$, it is natural to generalize the discussion and ask if M-theory on $AdS_4\times Y_7$ is dual to a certain Chern-Simons-matter theory. In order to preserve supersymmetry, the seven-dimensional manifold $Y_7$ should be either 3-Sasaki-Einstein, Sasaki-Einstein or weak $G_2$ manifold, and the corresponding gauge theory (in UV) should have $\mathcal{N}=3,2,1$ supersymmetry, respectively.

In this paper, we consider the case when $Y_7$ is a toric Sasaki-Einstein manifold \footnote{See \cite{JT} for a discussion in the case of toric hyperK\"ahler manifold. Since $\mathcal{N}=3$ is a special case of $\mathcal{N}=2$, our analysis should apply to their case as well. See also \cite{OP} for the discussion of squashed $S^7$, which is an example of weak $G_2$ manifold with corresponding UV gauge theory having $\mathcal{N}=1$.}. In this case, the corresponding Chern-Simons-matter theories have $\mathcal{N}=2$, and the vacuum moduli space was analyzed first in \cite{IK} and later more generally in \cite{MS}. Still, it is not known how to extract toric data of Calabi-Yau 4-fold $C(Y_7)$ (cone of $Y_7$) given the gauge theory data. In particular, it was not clear how to relate the level of Chern-Simons-matter theories to the coordinates of the lattice points of the toric diagram.
In principle, this should be possible since the vacuum moduli space should coincide with $C(Y_7)$.

%There it was found that the moduli space of a three-dimensional Chern-Simons-matter theory specified by a toric diagram is in close connection with that of a four-dimensional quiver gauge theory with the {\it same} quiver diagram. Namely, the conditions corresponding to F-flatness conditions are the same, but the D-flatness conditions are different, and is specified by levels of Chern-Simons terms.

The primary purpose of this paper is to propose a concrete algorithm to obtain the toric data of the toric Calabi-Yau 4-folds dual to three-dimensional Chern-Simons-matter quiver gauge theories.
 Our method applies to arbitrary toric Calabi-Yau 4-folds, and utilizes in an essential way some of the techniques from dimer models (brane tilings). 

The organization of the present article is as follows. First, in section \ref{sec.moduli}, we briefly review the vacuum moduli space of $\mathcal{N}=2$ Chern-Simons-matter theories, following \cite{IK,MS,HZ}. In the next section (section \ref{main.sec}) we briefly review dimer techniques, and then present our main result, namely the method to extract the toric data of the Calabi-Yau 4-fold from gauge theory data (the bipartite graph and the levels of Chern-Simons). In section \ref{eg.sec} we illustrate our method further by studying more examples.
The final section is devoted to summary and discussions.

%%%%%%%%%%%%%%%%%%%%%%%%%%%%%%%%%%%%%%%%%%%%%%%%%%%%%%%%%%%%%%%%%%%%%%%%
%%%%%%%%%%%%%%%%%%%%%%%%%%%%%%%%%%%%%%%%%%%%%%%%%%%%%%%%%%%%%%%%%%%%%%%%
\section{Moduli spaces of Chern-Simons-matter theories}\label{sec.moduli}
Let us begin by recalling some facts about $\mathcal{N}=2$ superconformal Chern-Simons-matter theories specified from a toric diagram. We will concentrate on the description of vacuum moduli space (VMS), which is discussed first by \cite{IK} and later more generally by \cite{MS,HZ}. Here we mostly follow \cite{MS}. In particular, it is expected that (for class of Chern-Simons-matter theories which have gravity duals) the VMS of the Chern-Simons-matter theories dual to M-theory on $AdS_4\times Y_7$ (where $Y_7$ is a seven-dimensional Sasaki-Einstein manifold) coincide with the Calabi-Yau 4-fold cone $C(Y_7)$.

We are going to study Chern-Simons-matter theories specified from a quiver diagram. Here a quiver diagram is simply an oriented graph, consisting of vertices (which we denote by $i,j,\ldots \in \scV$, where $\scV$ denotes the set of vertices) and oriented arrows (which we denote by $a,b,\ldots\in \scA$, where $\scA$ denotes the set of arrows). Each arrow $a$ has its source and target, which is denoted by $s(a)$ and $t(a)$, respectively. Just as in four-dimensional quiver gauge theories, vertices correspond to gauge groups, and arrows correspond to bifundamentals. Namely, we have a vector multiplet $V_i$ for each $i\in \scV$ and a chiral multiplet $\Phi_a$ for each $a\in \scA$. The gauge group is given by
\beq
G=\prod_{i=1}^n U(N_i),
\eeq
where $n$ is the number of vertices of the quiver diagram, and $N_i$ is the rank of the gauge group at vertex $i$. Since here we are considering Chern-Simons term in three dimensions, we have another parameters $k_i$ (the level of Chern-Simons term) for each vertex $i$. Also, in this paper we concentrate on the Abelian theories, namely $N_i=1$ for all $i$.

Given a quiver diagram, a superpotential and a set of parameters $N_i$'s and $k_i$'s, the corresponding $\scN=2$ Chern-Simons-matter theory is determined. For example, it is possible to write down the explicit form of the Lagrangian. 
However, since we want to focus on the vacuum moduli space (VMS) of the theory,  we here just write down the potential of the theory.

After deleting auxiliary fields, the potential $V$ is given by the sum of F-term potential and D-term potential:
\beq
V=V_F+V_D,
\eeq
where
\beq
V_F=\sum_{a\in \scA}\left|\frac{\partial W}{\partial \phi_a}\right|^2,
\eeq
and
\beq
V_D=\sum_{a\in \scA} |\phi_a|^2 (\sigma_{s(a)}-\sigma_{t(a)})^2.
\eeq
Here $\phi_a$ denotes the scalar component of the chiral multiplet $\Phi_a$ ($a\in \scA$), and $\sigma_i$ denotes the scalar in the vector multiplet $V_i$ ($i\in \scV$). Recall that three-dimensional $\scN=2$ vector multiplet is conveniently obtained by dimensionally reducing $\scN=1$ vector multiplet in four dimensions, and $\sigma_i$ is the component of the gauge field corresponding to the fourth direction. Also, we have a set of constraints, coming from the equations of motion of auxiliary fields of vector multiplets:
\beq
\scD_i=\frac{k_i \sigma_i}{2\pi},
\eeq
where $\scD_i$ is the usual 4d D-term
\beq
\scD_i=-\sum_{a|s(a)=i}|\phi_a|^2+\sum_{a|t(a)=i}|\phi_a|^2. \label{D=sigma}
\eeq 
Since nothing is charged under overall diagonal $U(1)$, it follows from \eqref{D=sigma} that
\beq
\sum_{i=1}^n k_i \sigma_i=0. \label{ksigma=0}
\eeq

In order to obtain the VMS, we have to minimize the potential, or make the potential vanish. The conditions coming from F-terms are simply
\beq
\frac{\partial W}{\partial \phi_a}=0,
\eeq
which defines a set
\beq
\scZ=\{dW=0\}\in \bC^{|\scA|}.
\eeq
This space $\scZ$ is sometimes called the ``master space'' and is recently studied in \cite{Master}.

We now turn to D-term equations.
 Interestingly, there are several types of solutions to D-term equations. In this paper, we concentrate on the particular set of solutions defined by 
\beq
\sigma_1=\sigma_2=\ldots =\sigma_n\equiv s, \label{sigma=s}
\eeq
where $s\in \bR$ is arbitrary, although other branches might be interesting. The conjecture by Martelli and Sparks \cite{MS} is that this is the branch of the moduli space which becomes a toric Calabi-Yau 4-fold cone $C(Y)$.  
Our following analysis strongly supports this conjecture.

In this branch, we have
\beq
\scD_i=\frac{k_i s}{2\pi},
\eeq
and \eqref{ksigma=0} becomes
\beq
\sum_{i=1}^n k_i=0. \label{sumk=0}
\eeq
This shows that \eqref{sumk=0} is a necessary condition for the branch \eqref{sigma=s} to exist. We will henceforce assume that \eqref{sumk=0} is satisfied.

%----------------------------------------

In the analysis for far, we have completely neglected gauge fields. Now what is special about three dimensions is that a gauge field is dualized (and thus equivalent) to a periodic scalar, and thus should be included when computing the VMS. Of course, basically we can forget about gauge fields since gauge fields are gauged away by gauge transformations. However, this still leaves constant gauge transformations, and after some analysis \cite{IK,MS} the result is the following.

Consider a character
\beq
\chi: & &U(1)^n\to U(1)\\
& &(e^{i\theta_1},\ldots, e^{i\theta_n})\mapsto \exp(i\sum_{i=1}^n k_i\theta_i).
\eeq
Define a group $G_{3d}$ by
\beq
G_{3d}=\textrm{ker}\chi / U(1),
\eeq
where $U(1)$ is the overall diagonal $U(1)$, which is clearly in $\mathrm{ker}\chi$ by \eqref{sumk=0}. Then this group $G_{3d}$ acts as the (effectively acting) group of gauge transformations, and the moduli space of 3d theory is given by
\beq
\scM_{3d}=\scZ//G_{3d}.
\eeq

This moduli space has an interesting connection to the moduli space of 4d theory \cite{MS,JT}. If we divide by the whole $G_{4d}\equiv U(1)^n/U(1)$ (here we are again removing overall $U(1)$ which acts trivially on all fields), then we have the moduli space of 4d theory:
\beq
\scM_{4d}=\scZ//G_{4d}.
\eeq 
In order words, we have the relation
\beq
\scM_{4d}=\scM_{3d}//U(1),
\eeq
where the $U(1)$ in the last expression is generated by an element of $G_{4d}\equiv U(1)^n/U(1)$ which does not belong to $G_{3d}=\mathrm{ker}\chi/U(1)$. More precisely, $\scM_{4d}$ is (for fixed $s\ne 0$) part of the baryonic branch of moduli space of 4d theory (see \cite{MSbaryon} for recent discussions). 

% We later concentrate on the case where $\scM_{4d}$ is a toric Calabi-Yau 3-fold. In this case (assuming that $\scM_{3d}$ is connected?) $\scM_{4d}$ is a toric Calabi-Yau 4-fold. This is because for each $s\in \bR$, $\scM_{3d}(s)$ is a $U(1)$ fibration of $\scM_{4d}$, whose fiber vanishes at special value of $s\in \bR$.

We finish this section with a comment on non-Abelian VMS, although we will concentrate on the Abelian case in all other parts of this paper. If $N_i=N$ for all $i\in \scV$, it was shown \cite{MS} that VMS becomes the symmetric product of $N=1$ VMS:
\beq
\scM_{3d,N}=\textrm{Sym}^N \scM_{3d,1}.
\eeq
This is consistent with the interpretation that in the dual gravity picture $N$ M2-branes are probing the Calabi-Yau cone $C(Y_7)$.

\section{Toric Calabi-Yau 4-folds from Chern-Simons-matter theories}\label{main.sec}

Let us now move to our proposal. As said previously, if the 3d theory is really dual to M-theory on $AdS_4\times Y_7$, then the VMS of 3d theory $\scM_{3d}$ should coincide with Calabi-Yau 4-fold cone $C(Y_7)$. This is important, since it says that the possible candidate for gravity dual $Y_7$ of 3d theories can be determined in this way.

The problem we want to consider in this paper is to give an efficient procedure to obtain $C(Y_7)$ beginning with a quiver diagram, superpotential and CS levels ($k_i$'s). In this paper, we concentrate on the case where dual Calabi-Yau 4-fold $C(Y_7)$ is toric, which makes the analysis technically tractable but still contains infinitely many highly non-trivial examples.\footnote{The superpotential must satisfy a non-trivial conditions in order for $C(Y_7)$ to be toric. There are some discussions on this in $AdS_5/CFT_4$ context. For example, it is necessary that each bifundamental should appear exactly twice in the terms of the superpotential, which is sometimes called the toric condition \cite{Symmetries}.}

Now since toric Calabi-Yau 4-folds are specified by a toric polytope, the problem is to extract the toric data beginning with a quiver diagram, a superpotential and a set of levels ($k_i$'s). Of course, as we have stressed, it is in principle possible to solve this problem since $\scM_{3d}$ should coincide with $C(Y_7)$. However, it is desirable to have simple combinatorial procedure to carry this out. In fact, in the context of $AdS_5/CFT_4$ correspondence, this problem is dubbed the ``forward problem'' and we have now a well-established combinatorial procedure to do this, using techniques from dimer models, such as Kasteleyn matrices and height functions \cite{BT}.
We now show in the following that dimer model techniques are also useful for solving the problem at hand.

\subsection{Review of CY$_3$ case}\label{subsec.review}
Before describing our proposal, let us briefly review the relevant dimer model techniques. Interested readers are referred to reviews \cite{Kennaway,Y} for more details and related topics. For conventions such as orientation of arrows, we basically follow \cite{Y}.

\begin{figure}[htbp]
\centering{\includegraphics[scale=0.5]{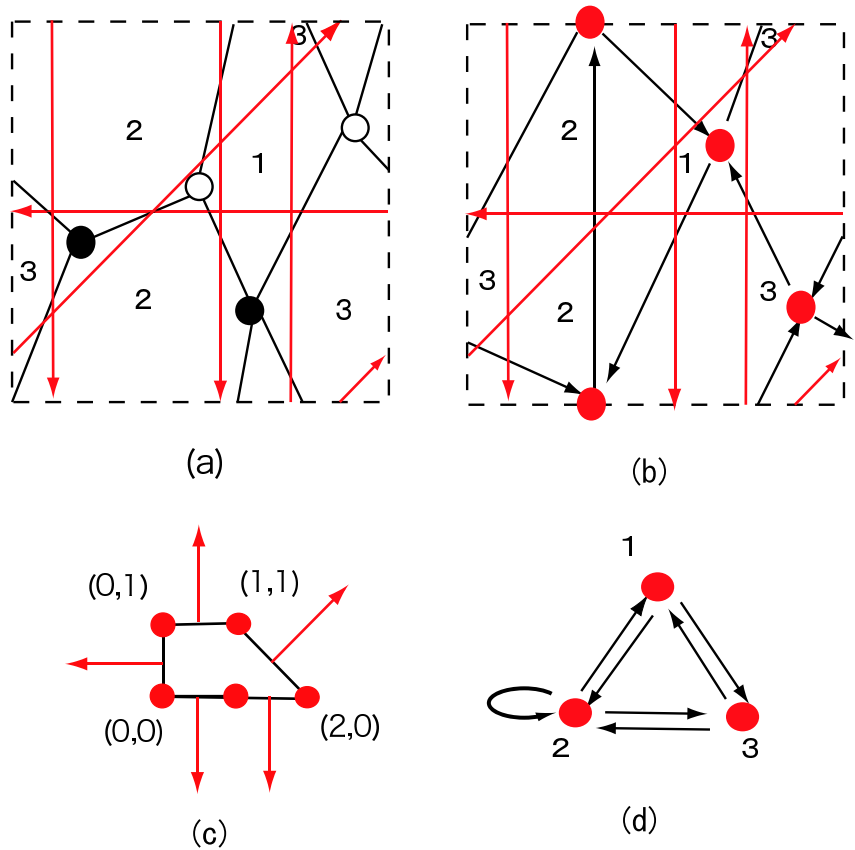}}
\caption{The bipartite graph (a), quiver diagram (b,d) and toric diagram (c) corresponding to the Suspended Pinched Point.}
\label{fig.SPP}
\end{figure}

Quiver gauge theories are usually specified by a quiver diagram, but a bipartite graph on $\bT^2$ (the so-called brane tiling) is far more powerful and thus we begin with a bipartite graph. A bipartite graph is a graph consistenting of vertices colored either black or white and vertices connecting different colors. Bipartite graphs we consider will always be written on two-dimensional torus $\bT^2$. See the bipartite graph of Figure \ref{fig.SPP} for an example, whose corresponding toric Calabi-Yau 3-fold is often called the Suspended Pinched Point (SPP) in the literature, and its corresponding quiver diagram and toric diagram (of the Calabi-Yau 3-fold $C(Y_5)$) is given in Figure \ref{fig.SPP}. The superpotential corresponds to a vertex in the bipartite graph, with $\pm$ sign determined according to the color of the vertex.

\begin{figure}[htbp]
\centering{\includegraphics[scale=0.3]{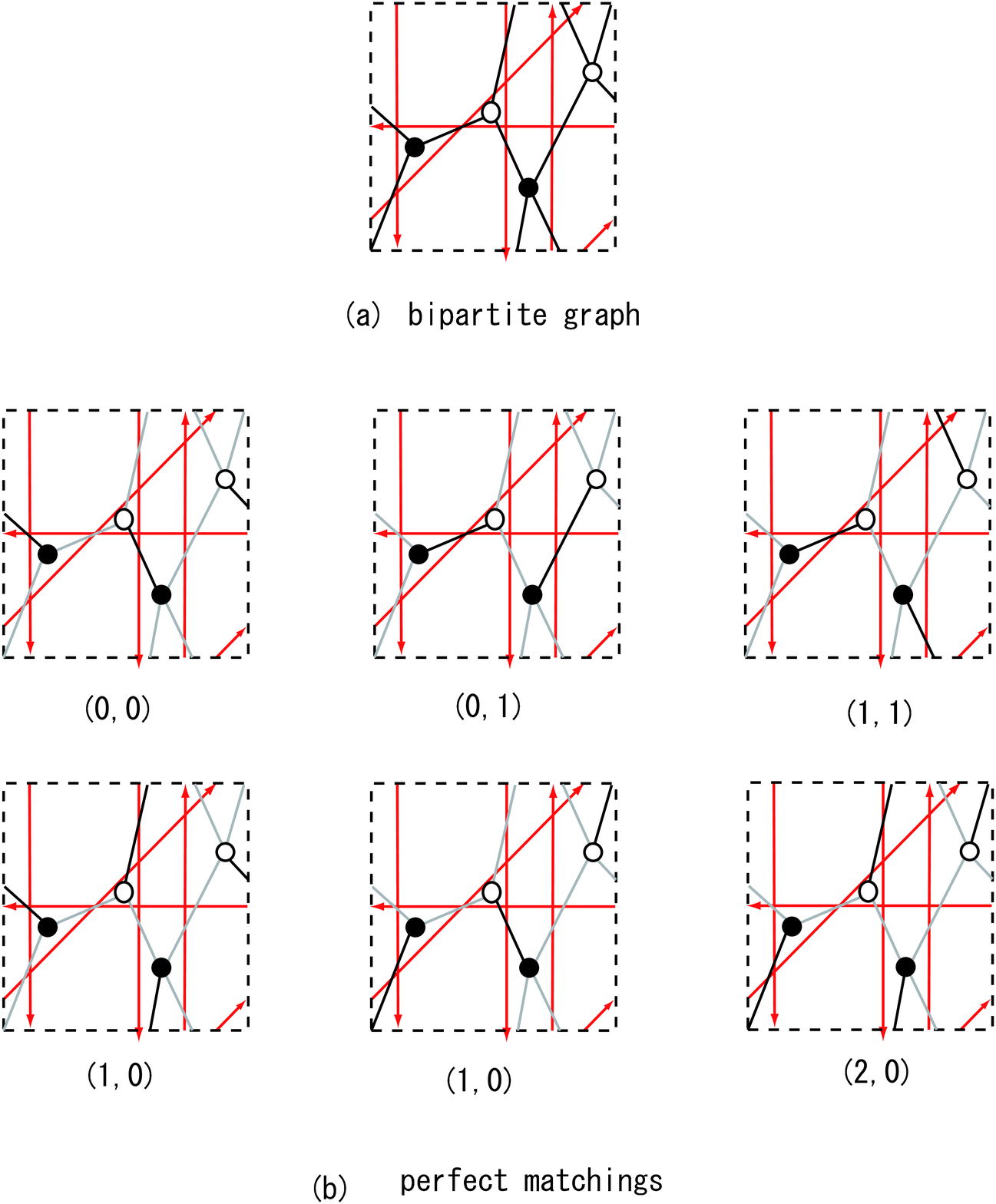}}
\caption{The perfect matchings of bipartite graphs and their height functions. The height functions gives the coordinates of the toric diagram.}
\label{fig.SPPfast}
\end{figure}

On a bipartite graph, we can define a perfect matching as a subset of edges which contains each vertex precisely once. In the case of SPP, we have 6 perfect matchings, as shown in Figure \ref{fig.SPPfast}. 

Now choose an arbitrary perfect matching $D_0$ and fix it as a reference matching. Then if we superimpose $D_0$ with another matching $D$, then we have a set of closed lines, which we denote by $D-D_0$. The height function is defined by the (signed) intersection number of $D-D_0$ with $\alpha$ and $\beta$ cycles of $\bT^2$:
\beq
(h_1(D-D_0),h_2(D-D_0))\equiv (\langle D-D_0, \alpha\rangle, \langle D-D_0, \beta\rangle ).
\label{eq.height}
\eeq
Then the intersting observation by \cite{BT} is that the convex hull of lattice points $(h_1(D-D_0), h_2(D-D_0))$ now coincides with the toric diagram $\Delta\subset \bZ^2$ (there are some ambiguities associated to the choice of $D_0$ and $\alpha, \beta$ cycles, but actually $\Delta$ is unique up to possible ambiguities of $GL(3,\bZ)$). This fact is now proven \cite{FV} and is often called the ``fast-forward problem''. The example of this procedure for SPP is shown in Figure \ref{fig.SPPfast}.

In a consistent bipartite graph, it was suggested in \cite{HV} that the perfect matching corresponding to a vertex of the toric diagram is unique, and we asume throughout this paper that this is the case. Then for each vertex $w_{\alpha}$ of the toric diagram $\Delta$, we can find the corresponding perfect matching, which is denoted by $D_{\alpha}$. Perfect matchings corresponding to the vertices of $\Delta$ are called vertex perfect matchings.

\subsection{Our proposal and its proof}

We now explain our proposal in detail. Our method gives the toric data of the Calabi-Yau 4-fold from gauge theory data, namely from the bipartite graph (which encodes the quiver diagram and the superpotential) and the levels of Chern-Simons terms.

First, we choose four products of paths (bifundamentals) $p_1, \dots, p_4$ on the bipartite graph as follows. Let $p_1$ ($p_2$) be a path on the torus corresponding to $\alpha$- ($\beta$)-cycles of $\bT^2$. Let $p_4$ %\footnote{We use here $p_4$ instead of $p_3$ for notational reasons which will become clear later.} 
be a closed path encircling the vertex of the bipartite graph, and thus corresponding to a trivial element of $H^1(\bT^2,\bZ)$. Then the operators $\scO_1,\scO_2,\scO_4$, which correspond to the product of corresponding bifundamentals along paths $p_1,p_2, p_4$, are invariant under $G_{4d}=U(1)^n/U(1)$ and thus gauge invariant. They are usually called mesonic operators. As the remaining path $p_3$, we take a path (or more preciesly product of paths in general) whose corresponding operator $\scO_3$ is invariant under $G_{3d}=\mathrm{ker}\chi/U(1)$, but not under $G_{4d}=U(1)^n/U(1)$ \footnote{Another way of expressing this fact is that in three dimensions we can attach 't Hooft operators to make them gauge invariant. For example, in the case of ABJM theory \cite{ABJM}, $p_3$ gives the operator $C^k$ discussed in section 2.4 of \cite{ABJM}.}. They are charged under baryonic symmetries and thus should be called baryonic operators, although they are not gauge invariant in the usual 4d sense .

\begin{figure}[htbp]
\centering{\includegraphics[scale=0.5]{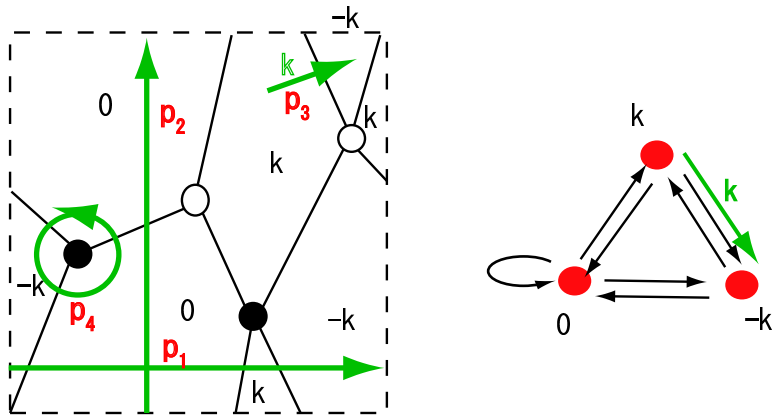}}
\caption{The choices of paths $p_1,\dots, p_4$ (left) and Chern-Simons levels for SPP (right). On the right figure the flow of the Chern-Simons charge is represented as a green arrow.}
\label{fig.SPPpaths}
\end{figure}

Now let us illustrate this method with the example shown in Figure \ref{fig.SPP}. We take the levels of Chern-Simons as shown in Figure \ref{fig.SPPpaths}. Then $\chi$ is given by 
\beq
(e^{i\theta_1},e^{i\theta_2}, e^{i\theta_3})\mapsto e^{i k (\theta_1-\theta_3)},
\eeq
and thus $\mathrm{ker}\chi$ is generated by $(e^{i\theta}, 1, e^{i\theta})$, $(1,e^{i\theta},1)$ and discrete subgroup $\bZ_k$.
In the figure the products of bifundamentals corresponding to $p_1,p_2,p_4$ are shown as closed paths, and $p_3$ is the $k$-th power of the bifundamental shown in the figure. It is easy to check that $p_3$ is indeed invariant under $\mathrm{ker}\chi$.
Of course, the choice of path is far from unique, and for example there is in general several choices of $\alpha$- and $\beta$-cycles. We will see, however, that this ambiguity does not matter in the final result. \footnote{More generally, we take $p_1,\dots p_4$ to be a minimal generating set of operators which are invariant under $\textrm{ker}\chi$.}

% There are some remarks about the choices of these paths. First remark is that they are just 
%  formal products of paths \footnote{Note for mathematically inclined readers: the word `product' is just a formal product, and different from the product in the path algebra of the quiver, which is defined by compositions of paths.}. Namely, they do not have to constitite a single path.

As you can see from this example, the path $p_4$ corresponds to a flow of  ``charges'' of levels of Chern-Simons terms 
(This point is discussed in \cite{IKnew}). Namely, we have a source of $k_i$ units of Chern-Simons levels at the $i$th face ($i$th gauge group), and the condition that $p_4$ is invariant under $\textrm{ker}\chi$ is precisely equivalent to the condition that this flow conserves the charge. This often makes the choice of $p_4$ easier. In the example of SPP, $p_4$ corresponds to a flow of charge as shown in Figure \ref{fig.SPPpaths}.

Now let $D_{\alpha}$ be the vertex perfect matching corresponding to a lattice point $w_{\alpha}$ of the toric diagram $\Delta$ and let $d_{\alpha}$ denote the corresponding toric divisor. Let $v^{\alpha}_i \ (i=1,\dots, 4)$ be the intersection number of $p_i$ with $D_{\alpha}$:
\beq
v^{\alpha}_i=\langle  p_i,D_{\alpha}\rangle.
\eeq
In particular, since $p_4$ is a path encircling a vertex of the bipartite graph, and since $D_{\alpha}$ is a perfect matching, $v^{\alpha}_4$ is always equal to one.

Now our proposal is that the toric polytope of Calabi-Yau 4-fold is given by the convex hull of all lattice points $v^{\alpha}=(v^{\alpha}_1,\dots v^{\alpha}_4$). This in particular states that the ambiguity we have encountered in the choice of paths $p_1,\dots p_4$ does not matter  up to $GL(4,\bZ)$-transformation of the toric polytope. This completes the description of our algorithm to obtain toric data from gauge theory data.\\

%-----------------------------------------

Let us again go back to the example of  SPP.
 In this example, using the paths $p_i$'s shown in Figure \ref{fig.SPPpaths} and perfect matchings shown in Figure \ref{fig.SPPfast}, the spanning vectors of the fan of $C(Y_7)$ are determined to be 
 \begin{equation}
 \begin{split}
 v_1&=(0,0,0,1),\quad v_2=(1,0,0,1),\quad  v_3=(2,0,0,1),  \\
 v_4&=(0,1,0,1), \quad v_5=(1,1,k,1).
 \end{split}
 \end{equation}

As a Calabi-Yau 4-fold, it corresponds to a certain toric orbifold of $\bC^4$, as discussed in \cite{IK} (see \cite{FTY,Benna,TY,Kim} for discussion of orbifolds of BLG models and ABJM models). For more general case of generalized conifolds,  completely parallel application of our method again yields the same result as in \cite{IK}.\\

%%------------------------------------------------

Our proposal seems ad hoc at first sight, but in fact we can give a proof of this proposal. To explain that we need two important facts.

First, in order to determine VMS we have to solve F-term equations. The convenient way to do this is to use perfect matchings. Namely, prepare a field $\rho_{\alpha}$ for each perfect matching $D_{\alpha}$, and then we can solve F-term equations by setting \cite{BT2}
\beq
\phi_a=\prod_{\alpha} \rho_{\alpha}^{\langle a, D_{\alpha}\rangle },
\label{eq.GLSM}
\eeq
 where $\phi_i$ is the scalar compoent of chiral superfield $\Phi_a$, and $\langle a, D_{\alpha}\rangle$ is the intersection number of arrow $a\in \scA$ (as written on $\bT^2$) and perfect matching $D_{\alpha}$. In other words, $\rho_{\alpha}$ are fields of Gauged Linear Sigma Model (GLSM) \cite{GLSM}.

The second important point is that toric divisors $d_{\alpha}$ (corresponding to $w_{\alpha}\in \Delta$) is described by an equation $\rho_{\alpha}=0$, where $\rho_{\alpha}$ is the GSLM field introduced in \eqref{eq.GLSM} (see Theorem 2 and Appendix A.2 of \cite{IKY} for the proof of this fact).  Namely, for each toric divisor $d$, we can associate a subset $D$ of $\scA$ of the set of edges of the dimer model so that $a \in D$ if and only if the arrows
crossed by $a$ is zero on the divisor $d$.
F-term relations coming from the superpotential implies
that $D$ contains a perfect matching,
and the condition that $d$ is a divisor
(i.e., it has the smallest possible codimension)
implies that $D$ is indeed a perfect matching.

Now the gauge invariant operators $\scO_1,\dots, \scO_4$ are products of $\phi_a$'s, and thus are expressed by the product of $\rho_{\alpha}$'s:
\beq
\scO_i=\prod_{\alpha}\rho_{\alpha}^{\langle p_i, D_{\alpha}\rangle}.
\eeq 
The gauge invariant operators should span the vacuum moduli space.
 Mathematically, each operator $\scO_i$ of paths defines
a $\bCx$-valued function on $\bT^4\subset \scM_{3d}$,
which can be extended to a $\bC$-valued fuction
on the moduli space.
By the general theory of toric variety (\cite{Fulton}, \S 3.3, Lemma), the $i$-th coordinate of the lattice point of the toric diagram corresponding to $D_{\alpha}$ is exactly the same as the order of zeros of $\scO_i$ at divisor $d_{\alpha}$. This completes our proof that $\langle p_i, D_{\alpha} \rangle$ gives the coordinate of the lattice point of the toric diagram. \footnote{Since perfect matchings are in one-to-one correspondence with terms in the determinant of Kasteleyn matrix, it automatically follows that we can rephrase our method by using Kasteleyn matrix with suitable weight.}

Note that by forgetting the path $p_3$, we are naturally lead back to the original story of ``fast-forward algorithm'' reviewed in section \ref{subsec.review}.
 Namely, $\langle p_1, D_{\alpha} \rangle \ (i=1,2)$ coincide with the height functions 
$h_i(D_{\alpha}) \ (i=1,2)$ as defined in \eqref{eq.height}.
Thus as a direct byproduct our analysis 
we have given a short alternative proof of the fast-forward algorithm, 
whose original proof is given in \cite{FV}.

\section{More examples}\label{eg.sec}
In the previous section, we have discussed the case where $Y_5$ is SPP, or more generally generalized conifolds. Although our techniques apply to arbitrary quivers and bipartete graphs corresponding to toric Calabi-Yau 3-folds $Y_5$, we study $Y^{p,k}(B_4)$ as a good illustrative example. 

In \cite{GMSW} it was shown that for any K\"ahler-Einstein manifold $B_4$ we can construct explicit metrics on the seven-dimensional Sasaki-Einstein manifold, which are denoted by $Y^{p,k}(B_4)$. Moreover, in version 2 of \cite{MS} it was conjectured that the quiver diagram for the Chern-Simons-matter theories exactly coincide with the quiver diagram corresponding to the Calabi-Yau 3-fold $K_{B_4}$, which is the canonical bundle over $B_4$.
 From the classification of K\"ahler-Einstein manifolds, $B_4$ is either 
 $\mathbb{CP}^1\times \mathbb{CP}^1, \mathbb{CP}^2$ or $dP_n$ (n=3,\dots, 8) \footnote{K\"ahler-Einstein metrics do not exist for $dP_1$ and $dP_2$ due to the existence of Matsushima obstruction.}, where $dP_n$ is the $n$ the del Pezzo surface obtained by blowing up $\mathbb{CP}^2$ at generic points $n$ times.
 Here we concentrate on the case of $B_4=\mathbb{CP}^1\times \mathbb{CP}^1$ and $B_4=\mathbb{CP}^2$.

\subsection{$Y^{p,k}(\mathbb{CP}^2)$}

\begin{figure}[htbp]
\centering{\includegraphics[scale=0.4]{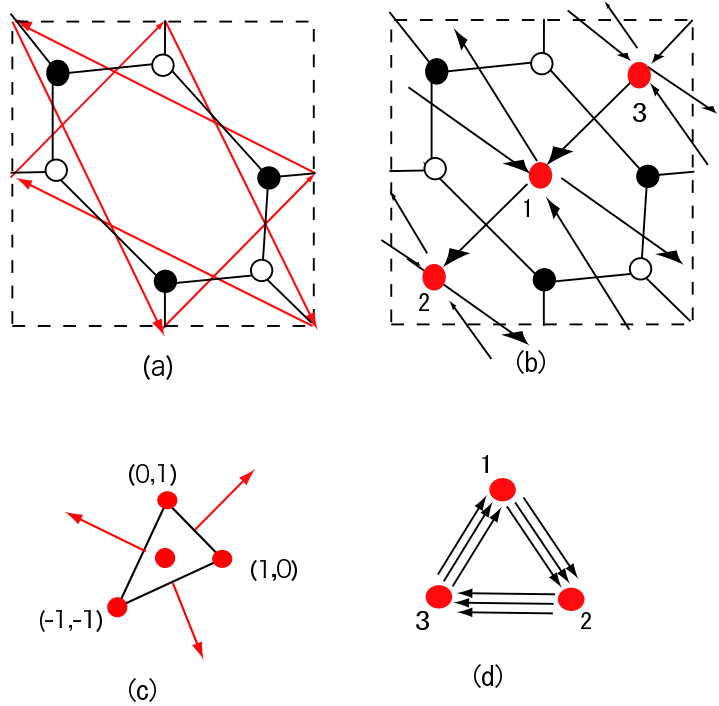}}
\caption{The bipartite graph (a), thee toric diagram (c) and the quiver diagram (b,d) for $C(Y_5)=K_{\mathbb{CP}^2}$.}
\label{fig.CP2}
\end{figure}

We begin with the case $B_4=\bC\bP^2$. The dual Chern-Simons-matter quiver is given in Figure \ref{fig.CP2}, which is obtained from the toric diagram of $\mathbb{CP}^2$ by standard dimer techniques (see also the recent discussion \cite{HZ}).

The levels of Chern-Simons terms are denoted by $k_1,k_2, k_3$, which sum up to zero by \eqref{sumk=0}. 
%For simplicity, we assume that $\textrm{gcd}(k_1,k_2)=1$. 
In this case, it is not difficult to find the paths $p_i$'s, and the final result is that the fan of $C(Y_7)$ is spanned by
\begin{equation}
\begin{split}
v_1&=(0,1,0,1), \quad v_2=(1,0,0,1), v_3=(-1,-1,k_1-k_3,1),\\
v_4&=(0,0,k_1,1), \quad v_5=(0,0,0,1), \quad v_6=(0,0,-k_3,1).
\end{split}
\end{equation}
If you set $k_1=-p, k_2=2p-k, k_3=-(k_1+k_2)=k-p$, then the convex polytope $\scP$ spanned by $v_1,\dots, v_5$ exactly coincides with the toric data provided in \cite{MStoric}. Of course, we have another vector $v_6$, but the condition for $v_0$ to lie in $\scP$ yields the condition $p \le k \le 2p$, just as analyzed in version 2 of \cite{MS}.

\begin{figure}[htbp]
\centering{\includegraphics[scale=0.3]{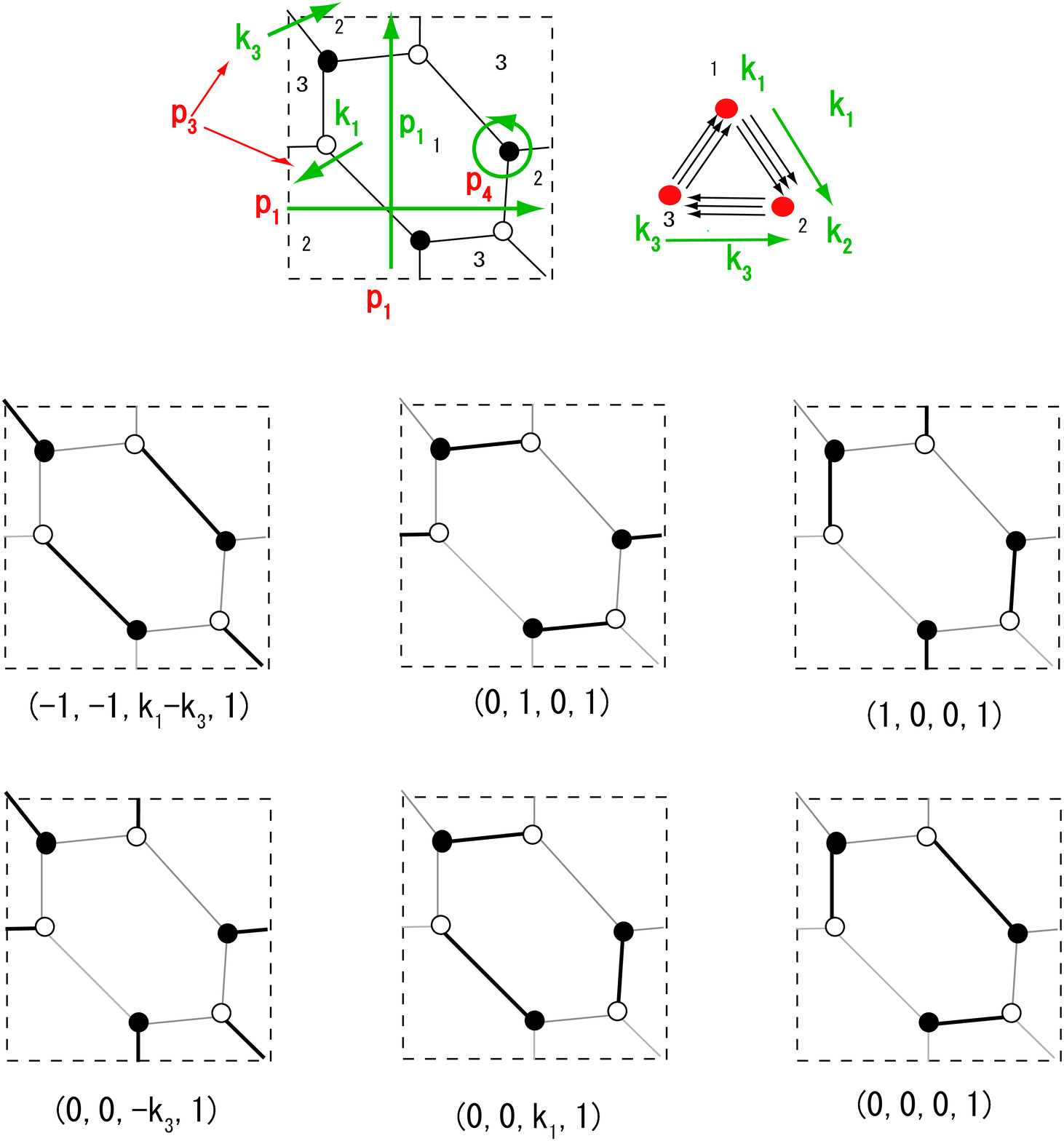}}
\caption{The choice of paths $p_1,\dots, p_4$, flow of CS cherge (above) and perfect matchings with corresponding coordinate of the lattice point of convex polytope (below) in the case of $C(Y_5)=K_{\mathbb{CP}^2}$.}
\label{fig.CP2PM}
\end{figure}

\subsection{$Y^{p,k}(\bC \bP^1\times \mathbb{CP}^1)$}
Now we treat the case of $B_4=\mathbb{CP}^1\times \mathbb{CP}^1$.
 In this case, there is no previous discussion of Chern-Simons-matter theories in the literature, but the procedure is completely analogous to the case of $B_4=\mathbb{CP}^2$.

\begin{figure}[htbp]
\centering{\includegraphics[scale=0.4]{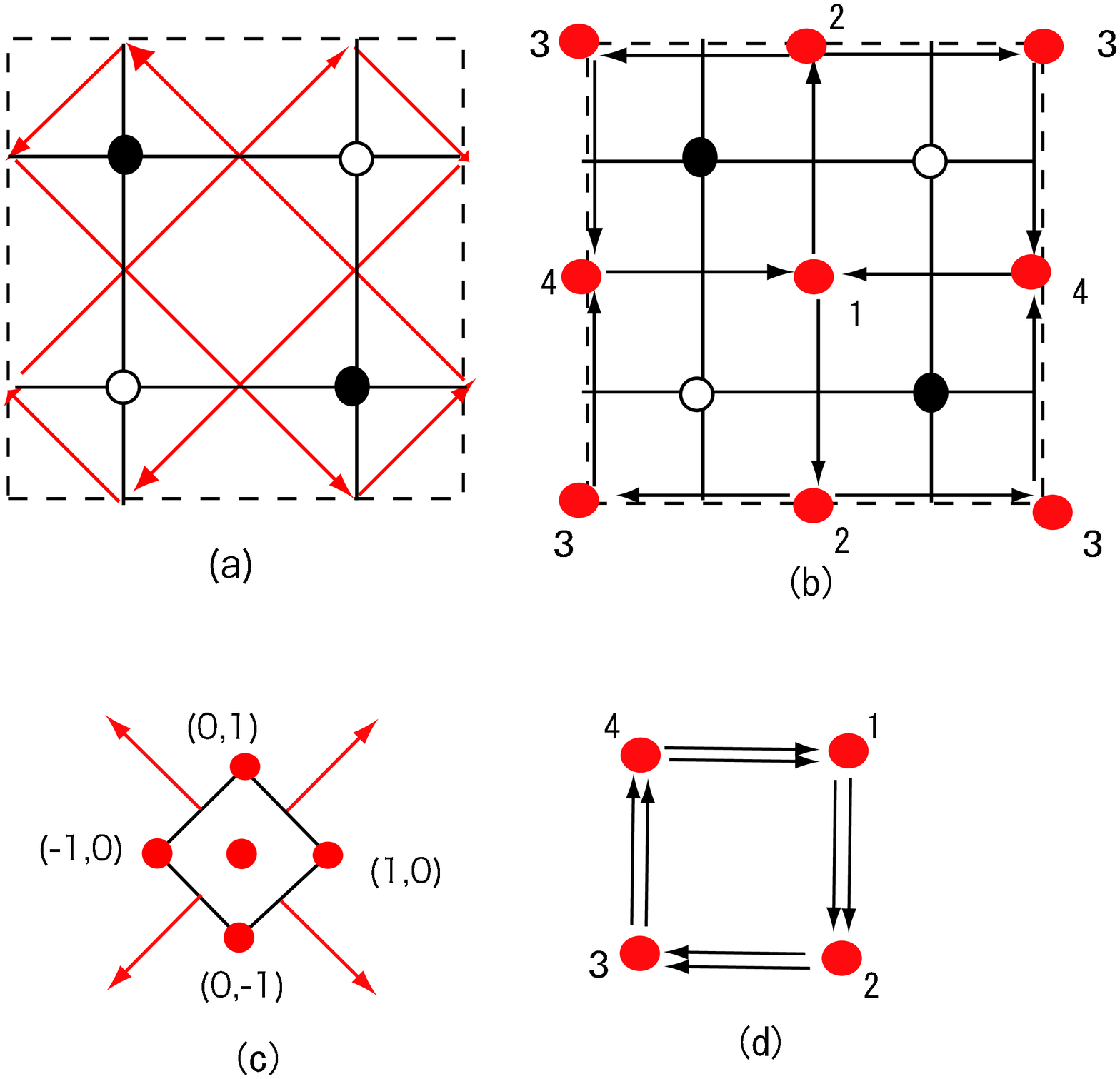}}
\caption{The bipartite graph (a), the toric diagram (c) and quiver diagrams (b,d) for $C(Y_5)=K_{\mathbb{CP}^1\times \mathbb{CP}^1}$.}
\label{fig.F0}
\end{figure}

\begin{figure}[htbp]
\centering{\includegraphics[scale=0.4]{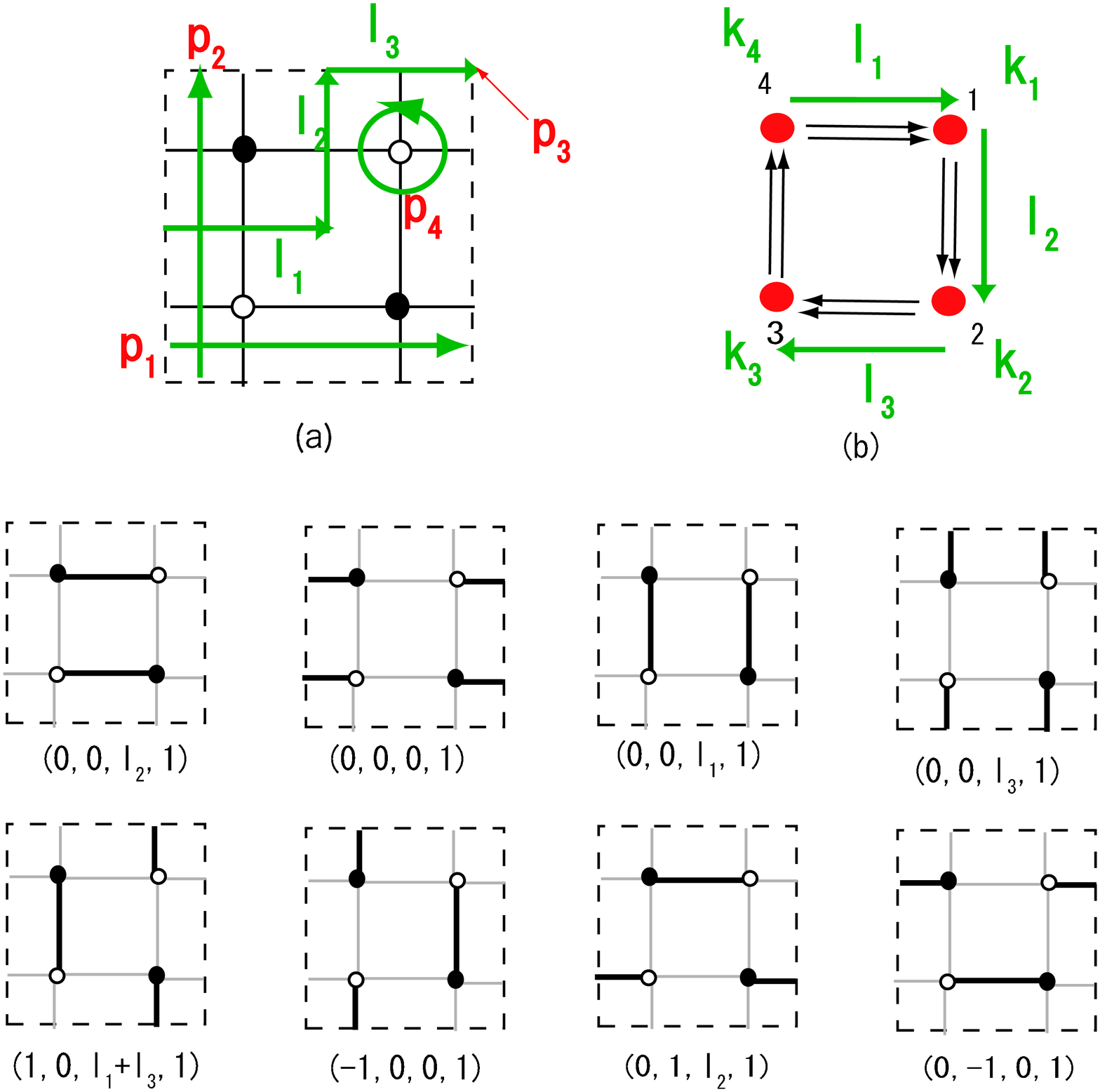}}
\caption{The perfect matchings and choice of paths $p_1,\dots, p_4$ for $C(Y_5)=K_{\mathbb{CP}^1\times \mathbb{CP}^1}$.}
\label{fig.F0PM}
\end{figure}

The bipartite graph and the quiver are shown in Figure \ref{fig.F0}, and the perfect matchings and the choice of paths are shown in Figure \ref{fig.F0PM}, where $l_i$'s ($i=1,2,3$) are determined by charge conservation, and thus $l_1=k_4, l_2=k_4+k_1, l_3=k_4+k_1+k_2=-k_3$. 
%We are now going to assume that $k_i$'s are all coprime to each other ($\textrm{gcd}(k_1,\dots, k_4)=1$).

From Figure \ref{fig.F0PM}, it is again a straightforward exercise to see that the fan of $C(Y_7)$ is spanned by
\begin{equation}
\begin{split}
v_1&=(0,0,0,1), \quad v_2=(0,0,l_1,1), \quad  v_3=(-1,0,0,1), \quad v_4=(1,0,l_1+l_3,1), \\
v_5&=(0,-1,0,1), \quad v_6=(0,1,l_2,1), \quad  v_7=(0,0,l_3,1), \quad v_8=(0,0,l_2,1).
\end{split}
\end{equation}
If we set $l_1=p, l_2=k, l_3=k-p$, namely $k_1=k-p, k_2=-p, k_3=p-k,k_4=p$, then 
the spanning vectors are given by
\begin{equation}
\begin{split}
v_1&=(0,0,0,1), \quad v_2=(0,0,p,1), \quad v_3=(-1,0,0,1), 
\quad v_4=(1,0,k,1), \\
 v_5&=(0,-1,0,1), \quad v_6=(0,1,k,1), \quad
v_7=(0,0,k-p,1), \quad v_8=(0,0,k,1).
\end{split}
\end{equation}
The first six vectors span a convex polytope $\scP$ which exactly matches with that of \cite{MStoric}. The condition for $v_7$ to be in $\scP$ is given by
\beq
p\le k\le 2p,
\eeq
which again is the same as the condition given by version 2 of \cite{MS}. However, we have another vector $v_8$, which is outside $\scP$. \\

\begin{figure}[htbp]
\centering{\includegraphics[scale=0.4]{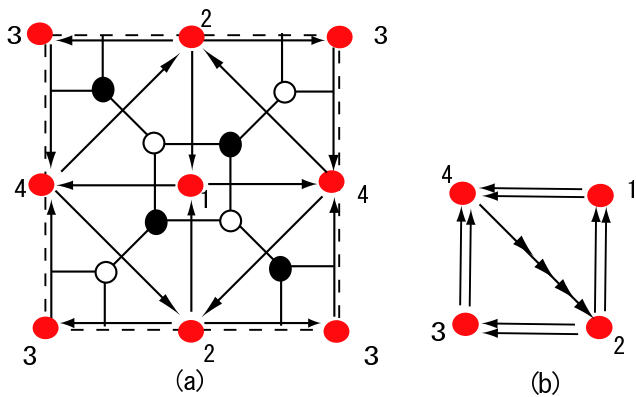}}
\caption{Another bipartite graph (a) and quiver diagram (b) for $C(Y_5)=K_{\mathbb{CP}^1\times \mathbb{CP}^1}$.}
\label{fig.F0phase2}
\end{figure}

%-----------------------------------------------------------------
This is still not the end of the story; for $\bP^1\times \bP^1$, we have another bipartite graph shown in Figure \ref{fig.F0phase2}. In the context of $AdS_5/CFT_4$  correspondence, the two bipartite graphs in Figures \ref{fig.F0} and \ref{fig.F0phase2} are related by the Seiberg duality \cite{BT2}. Since it is not clear whether our algorithm commutes with Seiberg duality, we will also study this another biparite graph explicitly.

\begin{figure}[htbp]
\centering{\includegraphics[scale=0.4]{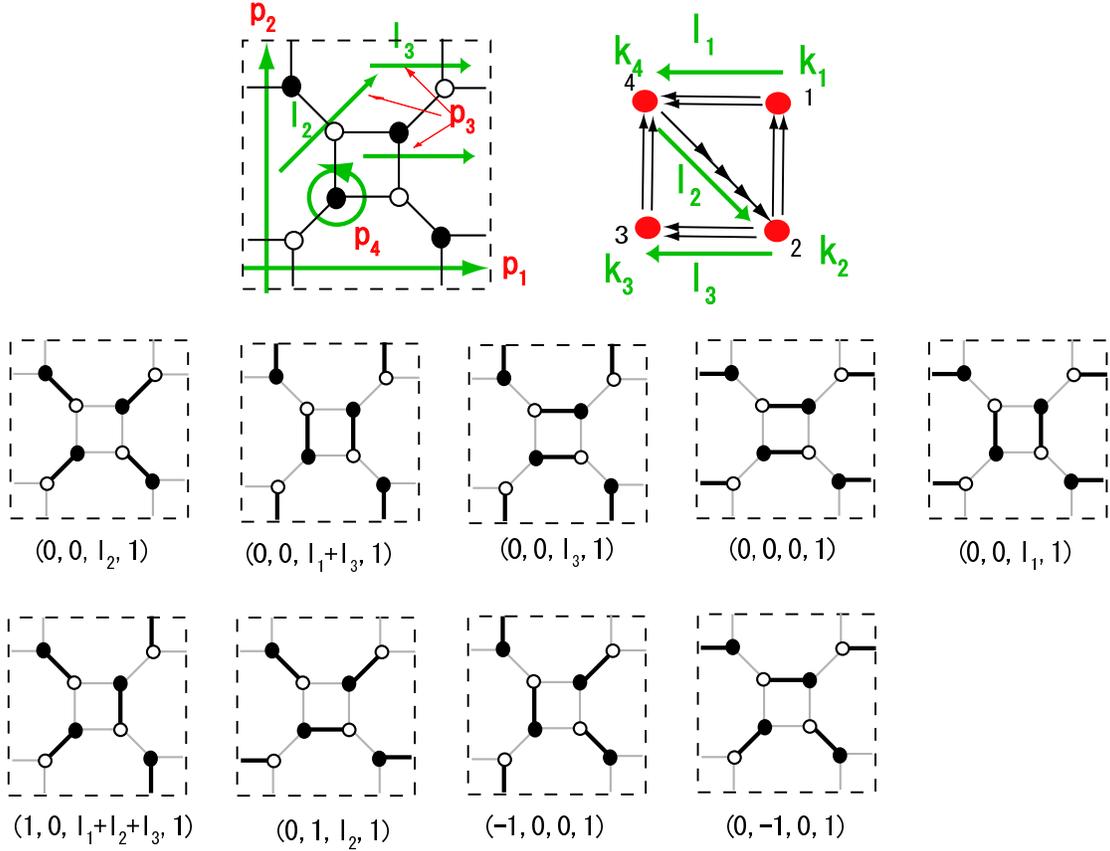}}
\caption{The perfect matchings and choice of paths $p_1,\dots, p_4$ for $C(Y_5)=K_{\mathbb{CP}^1\times \mathbb{CP}^1}$.}
\label{fig.F0phase2PM}
\end{figure}

The procedure to obtain toric data of $C(Y_7)$ is exactly the same as before. After taking appropriate paths $p_1,\ldots, p_4$ and listing all perfect matchings as in Figure \ref{fig.F0phase2PM} (from charge conservation we have $l_1=k_1, l_2=k_1+k_4, l_3=k_1+k_2+k_4=-k_3$), we see that the fan of $C(Y_7)$ is spanned by
\begin{equation}
\begin{split}
v_1&=(0,1,l_2,1), \quad  v_2=(-1,0,0,1), \quad v_3=(0,-1,0,1),\quad v_4=(0,0,0,1), \\
v_5&=(0,0,l_1,1), \quad  v_6=(0,0,l_3,1), \quad v_7=(1,0,l_1+l_2+l_3,1) \quad v_8=(0,0,l_2,1),\\
v_9&=(0,0,l_1+l_3,1).
\end{split}
\end{equation}

If we set $l_1=p, l_2=k, l_3=k-p$, namely $k_1=p, k_2=-p, k_3=-k+p,k_4=k-p$, then 
the spanning vectors are given by
\begin{equation}
\begin{split}
v_1&=(0,1,k,1), \quad  v_2=(-1,0,0,1), \quad v_3=(0,-1,0,1),\quad
v_4=(0,0,0,1),\\ v_5&=(0,0,p,1) \quad v_6=(0,0,k-p,1), \quad  v_7=(1,0,2k,1), \quad v_8=(0,0,k,1),\\
v_9&=(0,0,k,1).
\end{split}
\end{equation}
The first five vectors span a subset of convex polytope $\scP$ of version 2 of \cite{MS}, and the condition that $v_6$ is inside the polytope spanned by $v_1,\dots, v_5$ again gives $p \le k \le 2p$. However, in this case $v_7, v_8$ and $v_9$ are outside $\scP$. Thus we have shown that the conjecture by \cite{MS} does not hold for the case of $B_4=\mathbb{CP}^1\times \mathbb{CP}^1$. This also shows that our method of constructing toric data of $C(Y_7)$ does not commute with Seiberg duality in this example. It would be interesting to investigate this point further for more examples.

\section{Summary and discussions}
In this paper, we provided an explicit procedure to obtain toric data of a Calabi-Yau 4-fold dual to a large class of 3d Chern-Simons-matter theories specified by a bipartite graph (and thus a quiver diagram and a superpotential) and levels of Chern-Simons terms. We used techniques from dimer models, which was also crucial for the short proof of the correctness of our algorithm. We also analyzed the example of $C(Y_5)$ being given by generalized conifolds (SPP), $K_{\mathbb{CP}^2}$, $K_{\mathbb{CP}^1\times \mathbb{CP}^1}$. The analysis in the case of    $K_{\mathbb{CP}^1\times \mathbb{CP}^1}$ is inconsistent with the conjecture of \cite{MS}.

Of course, there are yet many issues to be discussed. For example, straightforward application of our techniques should yield infinitely many, rich set of examples of dual pairs of 3d Chern-Simons-matter theories and Calabi-Yau 4-folds. 
Using these results, it would be possible to study the issue of Seiberg(-like) duality \cite{Seiberg} for Chern-Simons matter theories (see \cite{GK} for recent discussions), or the inverse problem of obtaining bipartite graphs from the toric polytope of $C(Y_7)$ (see \cite{HV,FastInverse} for $AdS_5/CFT_4$ case).

At the same time, we should keep in mind that our procedure only gives the possible candidate for the dual gravity theory; even the existence of IR conformal fixed point of 3d theories is not clear in many cases and we are still far from verifying $AdS_4/CFT_3$ correspondence per se. In the case of $AdS_5/CFT_4$, we have powerful techniques to analyze 4d $\scN=1$ superconformal field theories, such as NSVZ $\beta$-function \cite{NSVZ} and a-maximization \cite{IW}, and it is desireble to find counterparts of these in $AdS_4/CFT_3$.

We have seen that brane tiling techniques are useful to the $AdS_4/CFT_3$ correspondence as well. However, the real meaning of dimers in this context is not yet clear, which is in constrast with the fact that brane tilings have a physical meaning as a projection of shapes of D5/NS5 brane configurations\cite{FHKV,IKY,Y} in the case of $AdS_5/CFT_4$. The recent works on the higher-dimensional generalization of dimer models\cite{Mcrystal} might give a possible clue to this problem.

Of course, we can envisage many generalization and applications, such as generalization to orientifolds \cite{FHKPU,IKY}, fractional branes and cascading \cite{fractional}, non-toric Sasaki-Einstein manifolds \cite{BFZ}, marginal deformations \cite{IIKY}, application to homological mirror symmetry \cite{UY}, all of which are analyzed in the $AdS_5/CFT_4$ context. We hope to return to these topics in the future.

\section*{Acknowlegments}
M.~Y. would like to thank Keisuke Kimura and Yosuke Imamura for valuable discussions and sharing some of their results, 
which was submitted to arXiv after the version 1 of this paper \cite{IKnew}.
We thank the anonymous referee for pointing out that
 our 2+1 dimensional construction does not commute with Seiberg duality.
This work is supported by Grant-in-Aid for Young Scientists 
(No.18840029) (K.~U.), by JSPS fellowships for Youngs Scientists and by World Premier International Research Center InitiativeiWPI Initiative), MEXT, Japan.
(M.~Y.).


\begin{thebibliography}{999}
\parskip=-2pt

%--------- brane tilings -------------------------

\bibitem{BT}
  A.~Hanany and K.~D.~Kennaway,
  ``Dimer models and toric diagrams,''
  arXiv:hep-th/0503149.
  %%CITATION = HEP-TH/0503149;%%

\bibitem{BT2}
  S.~Franco, A.~Hanany, K.~D.~Kennaway, D.~Vegh and B.~Wecht,
  %``Brane dimers and quiver gauge theories,''
  JHEP {\bf 0601}, 096 (2006)
  [arXiv:hep-th/0504110].
  %%CITATION = JHEPA,0601,096;%%

\bibitem{BT3}
  S.~Franco, A.~Hanany, D.~Martelli, J.~Sparks, D.~Vegh and B.~Wecht,
  %``Gauge theories from toric geometry and brane tilings,''
  JHEP {\bf 0601}, 128 (2006)
  [arXiv:hep-th/0505211].
  %%CITATION = JHEPA,0601,128;%%

\bibitem{equiv}
  M.~Bertolini, F.~Bigazzi and A.~L.~Cotrone,
  %``New checks and subtleties for AdS/CFT and a-maximization,''
  JHEP {\bf 0412}, 024 (2004)
  [arXiv:hep-th/0411249];
  %%CITATION = JHEPA,0412,024;%%

%\cite{Benvenuti:2004dy}
  S.~Benvenuti, S.~Franco, A.~Hanany, D.~Martelli and J.~Sparks,
  %``An infinite family of superconformal quiver gauge theories with
  %Sasaki-Einstein duals,''
  JHEP {\bf 0506}, 064 (2005)
  [arXiv:hep-th/0411264];
  %%CITATION = JHEPA,0506,064;%%


%\bibitem{BZ}
  A.~Butti and A.~Zaffaroni,
  %``R-charges from toric diagrams and the equivalence of a-maximization and
  %Z-minimization,''
  JHEP {\bf 0511}, 019 (2005)
  [arXiv:hep-th/0506232].
  %%CITATION = JHEPA,0511,019;%%


%--------- check of AdS/CFT------------------

%\cite{Intriligator:2003jj}
\bibitem{IW}
  K.~Intriligator and B.~Wecht,
  %``The exact superconformal R-symmetry maximizes a,''
  Nucl.\ Phys.\  B {\bf 667}, 183 (2003)
  [arXiv:hep-th/0304128].
  %%CITATION = NUPHA,B667,183;%%

\bibitem{MSY}
  D.~Martelli, J.~Sparks and S.~T.~Yau,
  %``The geometric dual of a-maximisation for toric Sasaki-Einstein
  %manifolds,''
  Commun.\ Math.\ Phys.\  {\bf 268}, 39 (2006)
  [arXiv:hep-th/0503183];
  %%CITATION = CMPHA,268,39;%%

  %``Sasaki-Einstein manifolds and volume minimisation,''
  Commun.\ Math.\ Phys.\  {\bf 280}, 611 (2008)
  [arXiv:hep-th/0603021].
  %%CITATION = CMPHA,280,611;%%
  
  
%\cite{Kachru:2003aw}
\bibitem{KKLT}
  S.~Kachru, R.~Kallosh, A.~Linde and S.~P.~Trivedi,
  %``De Sitter vacua in string theory,''
  Phys.\ Rev.\  D {\bf 68}, 046005 (2003)
  [arXiv:hep-th/0301240].
  %%CITATION = PHRVA,D68,046005;%%

%--------- "BGL theory"----------------

%\bibitem{Bagger:2006sk}
\bibitem{BLG}
  J.~Bagger and N.~Lambert,
  %``Modeling multiple M2's,''
  Phys.\ Rev.\  D {\bf 75}, 045020 (2007)
  [arXiv:hep-th/0611108];
  %%CITATION = PHRVA,D75,045020;%%

%  J.~Bagger and N.~Lambert,
  %``Gauge Symmetry and Supersymmetry of Multiple M2-Branes,''
  Phys.\ Rev.\  D {\bf 77}, 065008 (2008)
  [arXiv:0711.0955 [hep-th]];
  %%CITATION = PHRVA,D77,065008;%%
%\cite{Bagger:2006sk}

%  J.~Bagger and N.~Lambert,
  %``Comments On Multiple M2-branes,''
  JHEP {\bf 0802}, 105 (2008)
  [arXiv:0712.3738 [hep-th]].
  %%CITATION = JHEPA,0802,105;%%

%\cite{Gustavsson:2007vu}
  A.~Gustavsson,
  ``Algebraic structures on parallel M2-branes,''
  arXiv:0709.1260 [hep-th];
  %%CITATION = ARXIV:0709.1260;%%

%  A.~Gustavsson,
  ``Selfdual strings and loop space Nahm equations,''
  arXiv:0802.3456 [hep-th].
  %%CITATION = ARXIV:0802.3456;%%

%-----------------------------------------------

\bibitem{ABJM}
  O.~Aharony, O.~Bergman, D.~L.~Jafferis and J.~Maldacena,
  ``N=6 superconformal Chern-Simons-matter theories, M2-branes and their
  gravity duals,''
  arXiv:0806.1218 [hep-th].
  %%CITATION = ARXIV:0806.1218;%%

\bibitem{JT}
  D.~L.~Jafferis and A.~Tomasiello,
  ``A simple class of N=3 gauge/gravity duals,''
  arXiv:0808.0864 [hep-th].
  %%CITATION = ARXIV:0808.0864;%%

\bibitem{OP}
  H.~Ooguri and C.~S.~Park,
  ``Superconformal Chern-Simons Theories and the Squashed Seven Sphere,''
  arXiv:0808.0500 [hep-th].
  %%CITATION = ARXIV:0808.0500;%%

%----------------------------------------
\bibitem{IK}
  Y.~Imamura and K.~Kimura,
  ``Coulomb branch of generalized ABJM models,''
  arXiv:0806.3727 [hep-th].
  %%CITATION = ARXIV:0806.3727;%%

\bibitem{MS}
  D.~Martelli and J.~Sparks,
  ``Moduli spaces of Chern-Simons quiver gauge theories,''
  arXiv:0808.0912 [hep-th].
  %%CITATION = ARXIV:0808.0912;%%


\bibitem{HZ}
  A.~Hanany and A.~Zaffaroni,
  ``Tilings, Chern-Simons Theories and M2 Branes,''
  arXiv:0808.1244 [hep-th].
  %%CITATION = ARXIV:0808.1244;%%

%------------
\bibitem{Master}
  D.~Forcella, A.~Hanany, Y.~H.~He and A.~Zaffaroni,
  %``The Master Space of N=1 Gauge Theories,''
  JHEP {\bf 0808}, 012 (2008)
  [arXiv:0801.1585 [hep-th]].
  %%CITATION = JHEPA,0808,012;%%

\bibitem{MSbaryon}
  D.~Martelli and J.~Sparks,
  ``Symmetry-breaking vacua and baryon condensates in AdS/CFT,''
  arXiv:0804.3999 [hep-th].
  %%CITATION = ARXIV:0804.3999;%%

\bibitem{Symmetries}
  B.~Feng, S.~Franco, A.~Hanany and Y.~H.~He,
  %``Symmetries of toric duality,''
  JHEP {\bf 0212}, 076 (2002)
  [arXiv:hep-th/0205144].
  %%CITATION = JHEPA,0212,076;%%

%---- reviews----------------------------------

\bibitem{Kennaway}
  K.~D.~Kennaway,
  %``Brane Tilings,''
  Int.\ J.\ Mod.\ Phys.\  A {\bf 22}, 2977 (2007)
  [arXiv:0706.1660 [hep-th]].
  %%CITATION = IMPAE,A22,2977;%%

\bibitem{Y}
  M.~Yamazaki,
  %``Brane Tilings and Their Applications,''
  Fortschr.\ Phys.\ {\bf 56}, 555 (2008)
  [arXiv:0803.4474 [hep-th]].
  %%CITATION = ARXIV:0803.4474;%%


\bibitem{FV}
  S.~Franco and D.~Vegh,
  %``Moduli spaces of gauge theories from dimer models: Proof of the
  %correspondence,''
  JHEP {\bf 0611}, 054 (2006)
  [arXiv:hep-th/0601063].
  %%CITATION = JHEPA,0611,054;%%
  
  
  \bibitem{HV}
  A.~Hanany and D.~Vegh,
  %``Quivers, tilings, branes and rhombi,''
  JHEP {\bf 0710}, 029 (2007)
  [arXiv:hep-th/0511063].
  %%CITATION = JHEPA,0710,029;%%
  
  \bibitem{IKnew}
   Y.~Imamura and K.~Kimura,
   ``Quiver Chern-Simons theories and crystals,''
   arXiv:0808.4155 [hep-th].
   %%CITATION = ARXIV:0808.4155;%%

%--------------orbifold-----------------------
\bibitem{FTY}
  H.~Fuji, S.~Terashima and M.~Yamazaki,
  ``A New N=4 Membrane Action via Orbifold,''
  arXiv:0805.1997 [hep-th].
  %%CITATION = ARXIV:0805.1997;%%
\bibitem{Benna}
  M.~Benna, I.~Klebanov, T.~Klose and M.~Smedback,
  ``Superconformal Chern-Simons Theories and $AdS_4/CFT_3$ Correspondence,''
  arXiv:0806.1519 [hep-th].
  %%CITATION = ARXIV:0806.1519;%%
\bibitem{TY}
  S.~Terashima and F.~Yagi,
  ``Orbifolding the Membrane Action,''
  arXiv:0807.0368 [hep-th].
  %%CITATION = ARXIV:0807.0368;%%

\bibitem{Kim}
  N.~Kim,
  ``How to put the Bagger-Lambert theory on an orbifold : A derivation of the
  ABJM model,''
  arXiv:0807.1349 [hep-th].
  %%CITATION = ARXIV:0807.1349;%%
  
  
\bibitem{GLSM}
  E.~Witten,
  %``Phases of N = 2 theories in two dimensions,''
  Nucl.\ Phys.\  B {\bf 403}, 159 (1993)
  [arXiv:hep-th/9301042].
  %%CITATION = NUPHA,B403,159;%%

\bibitem{IKY}
  Y.~Imamura, K.~Kimura and M.~Yamazaki,
  %``Anomalies and O-plane charges in orientifolded brane tilings,''
  JHEP {\bf 0803}, 058 (2008)
  [arXiv:0801.3528 [hep-th]].
  %%CITATION = JHEPA,0803,058;%
  
\bibitem{Fulton}
  W.~Fulton, ``Introduction to toric varieties,"  
  Princeton University Press, 1993.
  

\bibitem{GMSW}
  J.~P.~Gauntlett, D.~Martelli, J.~F.~Sparks and D.~Waldram,
  %``A new infinite class of Sasaki-Einstein manifolds,''
  Adv.\ Theor.\ Math.\ Phys.\  {\bf 8}, 987 (2006)
  [arXiv:hep-th/0403038].
  %%CITATION = 00203,8,987;%%


%-------- discussions --------------------

\bibitem{MStoric}
  D.~Martelli and J.~Sparks,
  ``Notes on toric Sasaki-Einstein seven-manifolds and $AdS_4/CFT_3$,''
  arXiv:0808.0904 [hep-th].
  %%CITATION = ARXIV:0808.0904;%%

\bibitem{Seiberg}
  N.~Seiberg,
  %``Electric - magnetic duality in supersymmetric nonAbelian gauge theories,''
  Nucl.\ Phys.\  B {\bf 435}, 129 (1995)
  [arXiv:hep-th/9411149].
  %%CITATION = NUPHA,B435,129;%%



\bibitem{GK}
  A.~Giveon and D.~Kutasov,
  ``Seiberg Duality in Chern-Simons Theory,''
  arXiv:0808.0360 [hep-th];
  %%CITATION = ARXIV:0808.0360;%%

  V.~Niarchos,
  ``Seiberg Duality in Chern-Simons Theories with Fundamental and Adjoint
  Matter,''
  arXiv:0808.2771 [hep-th].
  %%CITATION = ARXIV:0808.2771;%%

  
\bibitem{FastInverse}
  J.~Stienstra,
  ``Hypergeometric Systems in two Variables, Quivers, Dimers and Dessins
  d'Enfants,''
  arXiv:0711.0464 [math.AG];
  %%CITATION = ARXIV:0711.0464;%%  
  
  D.~R.~Gulotta,
  ``Properly ordered dimers, $R$-charges, and an efficient inverse algorithm,''
  arXiv:0807.3012 [hep-th].
  %%CITATION = ARXIV:0807.3012;%%

\bibitem{NSVZ}
  V.~A.~Novikov, M.~A.~Shifman, A.~I.~Vainshtein and V.~I.~Zakharov,
  %``Exact Gell-Mann-Low Function Of Supersymmetric Yang-Mills Theories From
  %Instanton Calculus,''
  Nucl.\ Phys.\  B {\bf 229}, 381 (1983);
  %%CITATION = NUPHA,B229,381;%%
%  V.~A.~Novikov, M.~A.~Shifman, A.~I.~Vainshtein and V.~I.~Zakharov,
  %``Supersymmetric instanton calculus: Gauge theories with matter,''
  Nucl.\ Phys.\  B {\bf 260}, 157 (1985)
  [Yad.\ Fiz.\  {\bf 42}, 1499 (1985)];
  %%CITATION = YAFIA,42,1499;%%

%  V.~A.~Novikov, M.~A.~Shifman, A.~I.~Vainshtein and V.~I.~Zakharov,
  %``Beta Function In Supersymmetric Gauge Theories: Instantons Versus
  %Traditional Approach,''
  Phys.\ Lett.\  B {\bf 166}, 329 (1986)
  [Sov.\ J.\ Nucl.\ Phys.\  {\bf 43}, 294.1986\ YAFIA,43,459 (1986\ YAFIA,43,459-464.1986)].
  %%CITATION = YAFIA,43,459;%%

\bibitem{FHKV}
  B.~Feng, Y.~H.~He, K.~D.~Kennaway and C.~Vafa,
  ``Dimer models from mirror symmetry and quivering amoebae,''
  arXiv:hep-th/0511287.
  %%CITATION = HEP-TH/0511287;%%

%----- Mcrystal----------------------

\bibitem{Mcrystal}
  S.~Lee,
  %``Superconformal field theories from crystal lattices,''
  Phys.\ Rev.\  D {\bf 75}, 101901 (2007)
  [arXiv:hep-th/0610204];
  
  S.~Lee, S.~Lee and J.~Park,
  %``Toric AdS(4)/CFT(3) duals and M-theory crystals,''
  JHEP {\bf 0705}, 004 (2007)
  [arXiv:hep-th/0702120];
  %%CITATION = JHEPA,0705,004;%%

  S.~Kim, S.~Lee, S.~Lee and J.~Park,
  %``Abelian Gauge Theory on M2-brane and Toric Duality,''
  Nucl.\ Phys.\  B {\bf 797}, 340 (2008)
  [arXiv:0705.3540 [hep-th]];
  %%CITATION = NUPHA,B797,340;%%

  %%CITATION = PHRVA,D75,101901;%%
  K.~Hosomichi, K.~M.~Lee, S.~Lee, S.~Lee and J.~Park,
  %``N=4 Superconformal Chern-Simons Theories with Hyper and Twisted Hyper
  %Multiplets,''
  JHEP {\bf 0807}, 091 (2008)
  [arXiv:0805.3662 [hep-th]];
  %%CITATION = JHEPA,0807,091;%%

\bibitem{FHKPU}
  S.~Franco, A.~Hanany, D.~Krefl, J.~Park, A.~M.~Uranga and D.~Vegh,
  %``Dimers and Orientifolds,''
  JHEP {\bf 0709}, 075 (2007)
  [arXiv:0707.0298 [hep-th]].
  %%CITATION = JHEPA,0709,075;%%

\bibitem{fractional}
  S.~Franco, A.~Hanany, F.~Saad and A.~M.~Uranga,
  %``Fractional branes and dynamical supersymmetry breaking,''
  JHEP {\bf 0601}, 011 (2006)
  [arXiv:hep-th/0505040].
  %%CITATION = JHEPA,0601,011;%%

\bibitem{BFZ}
  A.~Butti, D.~Forcella and A.~Zaffaroni,
  %``Deformations of conformal theories and non-toric quiver gauge theories,''
  JHEP {\bf 0702}, 081 (2007)
  [arXiv:hep-th/0607147].
  %%CITATION = JHEPA,0702,081;%%

\bibitem{IIKY}
  Y.~Imamura, H.~Isono, K.~Kimura and M.~Yamazaki,
  %``Exactly marginal deformations of quiver gauge theories as seen from   brane
  %tilings,''
  Prog.\ Theor.\ Phys.\  {\bf 117}, 923 (2007)
  [arXiv:hep-th/0702049].
  %%CITATION = PTPKA,117,923;%%
  
\bibitem{UY}
  K.~Ueda and M.~Yamazaki,
  ``A Note on Brane Tilings and McKay Quivers,''
  arXiv:math/0605780;
  %%CITATION = MATH/0605780;%%

  ``Homological mirror symmetry for toric orbifolds of toric del Pezzo
  surfaces,''
  arXiv:math/0703267;
  %%CITATION = MATH/0703267;%%

  ``Brane tilings for parallelograms with application to homological mirror
  symmetry,''
  arXiv:math/0606548.
  %%CITATION = MATH/0606548;%%

\end{thebibliography}
\end{document}